\documentclass[aps,prl,floatfix,superscriptaddress,twocolumn,showpacs]{revtex4-1}
\usepackage{float}
\usepackage{color}
\usepackage{comment}
\usepackage{siunitx}
\usepackage{soul}

\usepackage{graphicx}  
\usepackage{dcolumn}   
\usepackage{multirow}  
\usepackage{bm}        
\usepackage{amssymb}   
\usepackage{amsmath}   
\usepackage{epstopdf}
\usepackage{url}
\usepackage{hyperref}  
\usepackage{breakurl}
\hypersetup{breaklinks = true, colorlinks = true, citecolor = blue, linkcolor = blue, urlcolor = blue}
\newcommand{\deep}{^2\textrm{H}(\vec{e},e^{\prime}\vec{p})n}
\newcommand{\gevc}{\ifmmode {\textrm{GeV}/c} \else $\textrm{GeV}/c$\fi}
\newcommand{\hefour}{\ifmmode {{}^4\textrm{He}} \else ${}^4\textrm{He}$\fi}

\begin{document}

\newcommand*{\TLV}{School of Physics and Astronomy, Tel Aviv University, Tel Aviv 69978, Israel.}
\newcommand*{\MIT}{Massachusetts Institute of Technology, Cambridge, MA 02139.}
\newcommand*{\Mainz}{Institut f\"ur Kernphysik, Johannes Gutenberg-Universit\"at, 55099 Mainz, Germany.}
\newcommand*{\Hebrew}{Racah Institute of Physics, Hebrew University of Jerusalem, Jerusalem 91904, Israel.}
\newcommand*{\Rutgers}{Rutgers, The State University of New Jersey, Piscataway, NJ 08855, USA.}
\newcommand*{\JSI}{Jo\v{z}ef Stefan Institute, 1000 Ljubljana, Slovenia.}
\newcommand*{\UL}{Department of Physics, University of Ljubljana, 1000 Ljubljana, Slovenia.}
\newcommand*{\columb}{University of South Carolina, Columbia, South Carolina 29208, USA.}
\newcommand*{\zagreb}{Department of Physics, University of Zagreb, HR-10002 Zagreb, Croatia.}
\newcommand*{\nrc}{Department of Physics, NRCN, P.O. Box 9001, Beer-Sheva 84190, Israel.}

\title{Polarization-transfer measurement to a large-virtuality bound proton in the deuteron}

\author{I.~Yaron}\thanks{These authors contributed equally to this work.}\affiliation{\TLV}
\author{D.~Izraeli}\thanks{These authors contributed equally to this work.}\affiliation{\TLV}
\author{P.~Achenbach}\affiliation{\Mainz}
\author{H.~Arenh\"ovel}\affiliation{\Mainz}
\author{J.~Beri\v{c}i\v{c}}\affiliation{\JSI}
\author{R.~B\"ohm}\affiliation{\Mainz}
\author{D.~Bosnar}\affiliation{\zagreb}
\author{L.~Debenjak}\affiliation{\JSI}
\author{M.\,O.~Distler}\affiliation{\Mainz}
\author{A.~Esser}\affiliation{\Mainz}
\author{I.~Fri\v{s}\v{c}i\'{c}}\thanks{Present~address:~MIT-LNS,~Cambridge,~MA~02139,~USA.}\affiliation{\zagreb}
\author{R.~Gilman}\affiliation{\Rutgers}
\author{I.~Korover}\affiliation{\TLV}\affiliation{\nrc}
\author{J.~Lichtenstadt}\affiliation{\TLV}
\author{H.~Merkel}\affiliation{\Mainz}
\author{D.\,G.~Middleton}\affiliation{\Mainz}
\author{M.~Mihovilovi\v{c}}\affiliation{\Mainz}
\author{U.~M\"uller}\affiliation{\Mainz}
\author{E.~Piasetzky}\affiliation{\TLV}
\author{S.~\v{S}irca}\affiliation{\UL}\affiliation{\JSI}
\author{S.~Strauch}\affiliation{\columb}
\author{J.~Pochodzalla}\affiliation{\Mainz}
\author{G.~Ron}\affiliation{\Hebrew}
\author{B.\,S.~Schlimme}\affiliation{\Mainz}
\author{M.~Schoth}\affiliation{\Mainz}
\author{F.~Schulz}\affiliation{\Mainz}
\author{C.~Sfienti}\affiliation{\Mainz}
\author{M.~Thiel}\affiliation{\Mainz}
\author{A.~Tyukin}\affiliation{\Mainz}
\author{A.~Weber}\affiliation{\Mainz}
\collaboration{A1 Collaboration}

\date{\today}
\label{dead}

\begin{abstract}
Possible differences between free and bound protons may be observed in the ratio of
polarization-transfer components, $P'_x/P'_z$.
We report the measurement of $P'_x/P'_z$, in the $\deep$ reaction at low and high missing momenta. 
Observed increasing deviation of $P'_x/P'_z$ from that of a free proton as a function of the virtuality, similar to that observed in \hefour, indicates that the effect in nuclei is due to the virtuality of the knock-out proton and not due to the average nuclear density. 
The measured differences from calculations assuming free-proton form factors ($\sim10\%$), may indicate in-medium modifications.

\end{abstract}


\maketitle


The proton elastic electric and magnetic form factors (FFs) describe the distributions of electric charge and magnetization inside the proton and thus are intimately related to its internal structure~\cite{paper1}. 
Scattering polarized electrons off protons and simultaneous measurement of two polarization-transfer components allow one to determine the ratio of the electric 
to magnetic FF~\cite{Akh74}. 
This method eliminates many potential sources of systematic uncertainties and allows for high-precision measurements of this ratio.

In nuclei, the effects of the strong nuclear field on a bound nucleon are an interesting issue: Do bound nucleons have the same properties as free ones~\cite{paper3}? 
Measurement of the polarization of a proton that was knocked out of a nucleus by a polarized electron (quasi-elastic scattering) and the comparison to that of a free proton enable to probe possible changes in the FF ratio and may point to changes in the internal structure of a bound proton.
It is particularly interesting to measure such effects in the deuteron  which is the  most weakly bound nuclear system, and is frequently used as a `free-neutron' target. 
Changes in the FF of `off-the-mass-shell' nucleons may be relevant even in the deuteron.

We report here measurements of the transverse and longitudinal polarization-transfer components of the knocked out proton in the quasi-free $\deep$ reaction as a function of the proton missing-momentum ($p_{\textrm{miss}}$) compared to those of the free proton~\cite{paper4}. 
The data are also compared to calculations~\cite{paper5}, as well as to existing similar data from the $^{4}\textrm{He}(\vec{e},e^{\prime}\vec{p})^3\textrm{H}$ reaction~\cite{paper6,paper7}. The comparison of data from a loosely bound nucleus (deuteron) to the data from the high 
density \hefour\ was made as a function of the proton virtuality defined as:

\vspace{-5mm}
\begin{align}
\nu\equiv \Big(M_A - \sqrt{M^{2}_{A-1} + p^{2}_{\textrm{miss}}}\Big)^{2} - p^{2}_{\textrm{miss}} - M^{2}_{p},
\label {eq_nu}
\end{align}
where $M_A$, $M_{A-1}$ and $M_p$ are the target, residual nucleus, and proton masses, respectively
$p_{\textrm{miss}}$ is the missing-momentum in the reaction, $\vec{p}_{\textrm{miss}} = \vec{q} - \vec{p}_{p}$, $\vec{q}$ is the momentum transfer and $\vec{p}_{p}$ is the out-going proton momentum.
The good agreement between the deuteron and \hefour\ data, shown below,  indicates that the deviations from the free proton ratio
do not originate from the average nuclear density (which is vastly different in these nuclei) but rather from the `off shell' effects of the knocked out proton, as reflected by its virtuality.

In the elastic $^1\textrm{H}(\vec{e},e^{\prime}\vec{p})$ reaction there are two beam helicity dependent, non-vanishing polarization-transfer components: transverse ($P'_x$, perpendicular to the proton momentum in the scattering plane defined by the incident and scattered electron), and longitudinal ($P'_z$, along the proton momentum). 
In the one photon exchange approximation their ratio $(P'_x/P'_z)_H$ is directly related to the ratio of elastic electric $G^p_{E}(Q^2)$ to magnetic $G^p_M(Q^2)$ FF at a given four-momentum transfer $Q^2$~\cite{Akh74}:

\vspace{-5mm}
\begin{align}
\left(\frac{P'_x}{P'_z}\right)_{H} = -\frac{2M_p}{(E+E^{\prime})\tan(\theta_e/2)}\cdot\frac{G^p_E(Q^2)}{G^p_M(Q^2)},
\end{align}
where $E$ ($E^{\prime}$) is the incident (scattered) electron energy, and $\theta_e$ is the electron scattering angle.

\begin{figure}
\center\includegraphics[width=0.4\textwidth]{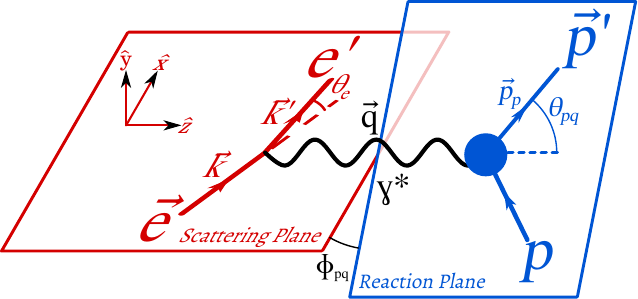}
\caption{The kinematics for quasi-elastic scattering of a bound proton in a nucleus, defining scattering and reaction planes.}
\label{PrlFig1} 
\end{figure}

For quasi-free elastic scattering off a bound nucleon, the knock-out proton is in general not ejected in the momentum-transfer ($\vec q$) direction (and in the scattering plane) due to its initial momentum. 
This introduces an additional plane, the reaction plane, spanned by the momentum transfer ($\vec q$) and the outgoing proton momentum ($\vec {p}_{p}$), characterized by the spherical angles $\theta_{pq}$ and $\phi_{pq}$ as shown in Fig.~\ref{PrlFig1}.
Following the convention of~\cite{paper7} the polarization components are the ones in the scattering plane along ($z$) and perpendicular ($x$) to $\vec q$.

Changes in the structure of a bound nucleon may be reflected in the measured ratio ($P'_{x}/P'_{z}$) of knocked-out nucleons. 
Other effects may influence this ratio as well.
We compare the measurement to state of the art calculations of the deuteron that take into account meson-exchange currents (MEC), isobar currents (IC), relativistic contributions of lowest order (RC), and final-state interactions (FSI)~\cite{paper5}. 
We observe significant differences between the calculation and the measured results.

The experiment was performed at the Mainz Microtron (MAMI) using the A1 beam-line and spectrometers~\cite{a1aparatus}. 
For the measurements \SIlist[list-units = single]{600;630}{MeV} CW polarized electron beams of \SI{10}{\micro \ampere} current were used. 
The average beam polarization was $80\%$, measured with a M{\o}ller polarimeter. 
The beam helicity was flipped at a rate of  \SI{1}{Hz}. The target consisted of an oblong shaped cell (\SI{50}{mm} long, \SI{11.5}{mm} diameter) filled with liquid deuterium. 
Two high-resolution, small solid angle, spectrometers, with momentum acceptances of $20-25\%$, were used to detect scattered electrons and knocked out protons, in coincidence.
The proton spectrometer was equipped with a focal-plane polarimeter (FPP) with a $3-7$ cm thick carbon analyzer~\cite{a1aparatus,Pospischil:2000pu}. 
The spin dependent scattering of the polarized proton by the carbon analyzer allows the determination of the proton longitudinal and transverse polarization components at 
the reaction point in the target~\cite{Pospischil:2000pu}. 
These polarization-transfer components were obtained by correcting for the spin precession in the spectrometer magnetic field.

The measurements were performed at four kinematic set-ups that covered two $Q^2$ ranges and two missing-momentum ranges each. 
Details of the kinematic settings are summarized in Table \ref{tab:Kinematics}, where $p_p$ and $\theta_p$  ($p_e$ and $\theta_e$) are the knock-out proton 
(scattered electron) momentum and angle.
The missing momentum is taken to be positive (negative) if a component of $\vec{p}_{\textrm{miss}}$ is parallel (anti-parallel) 
to the momentum-transfer vector ($\vec q$). 
\begin{figure}
\includegraphics[width=0.5\textwidth,height=2.6in]{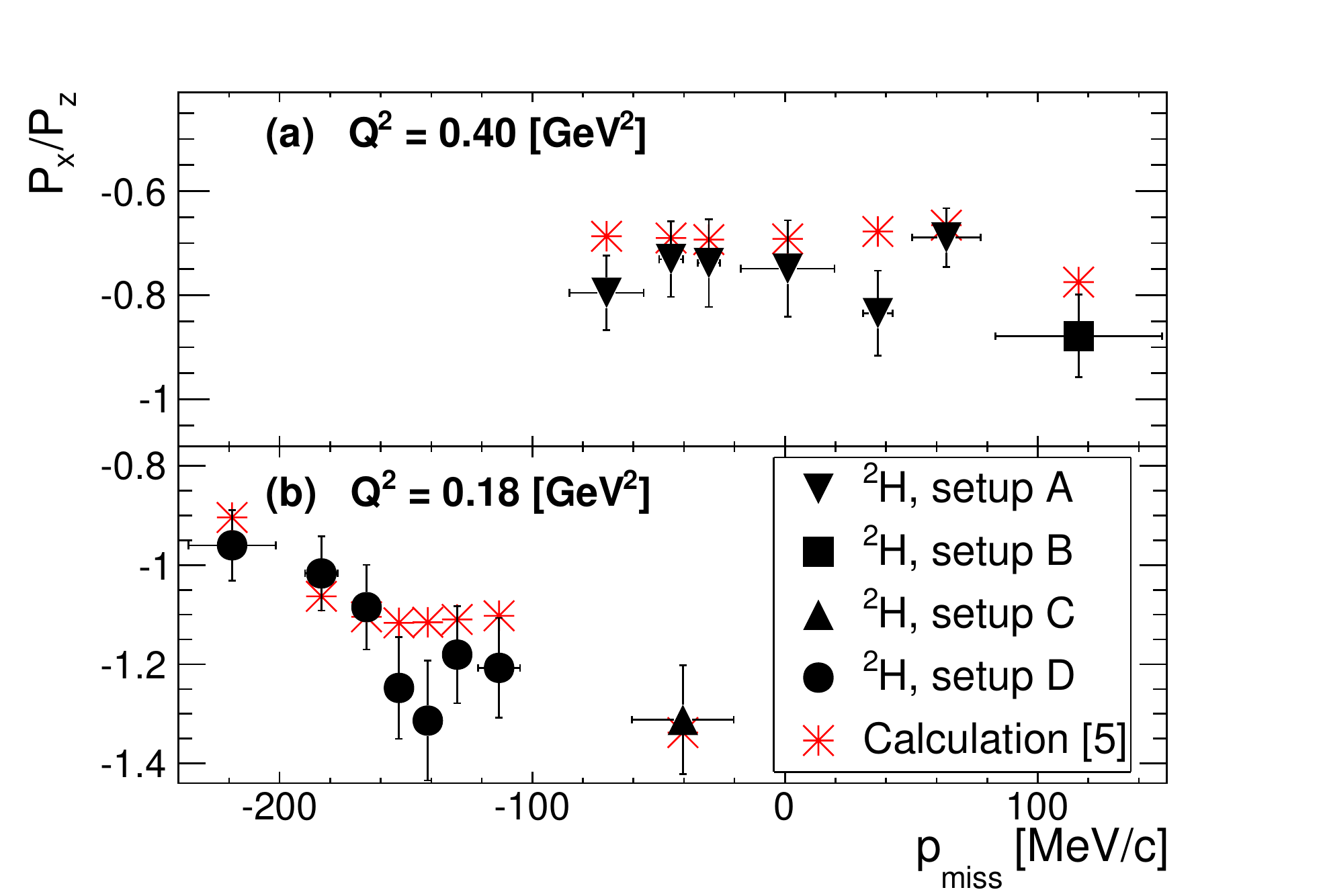}
\caption{\label{PrlFig2}The measured ratio of helicity dependent polarization components, $P'_x/P'_z$, versus the missing-momentum. The data are compared to the calculations based on the theoretical framework of Ref.~\cite{paper5}. 
The uncertainties are statistical only, and the horizontal bars indicate the $p_{\textrm{miss}}$ standard deviation in each bin.}
\end{figure}
\begin{table}[ht]
\caption{\label{tab:Kinematics}The kinematic settings in the experiment. 
The angles and momenta represent the center values for the two  spectrometer set-ups.}
\begin{ruledtabular}
\begin{tabular}{rlcccc}
\multicolumn{2}{c}{} & Setup & Setup & Setup & Setup\tabularnewline
\multicolumn{2}{c}{Kinematic} & A & B & C & D\tabularnewline
\hline 
$E_{\mathrm{beam}}$ & $\mbox{{[}MeV{]}}$ & 600 & 600 & 630 & 630\tabularnewline
$Q^{2}$ & $[\textrm{GeV}^{2}]$ & 0.40 & 0.40 & 0.18 & 0.18\tabularnewline
$p_{\mathrm{miss}}$ & $\mbox{{[}MeV{]}}$ &-80 to 75 & 75 to 175&-80 to -15 & -220 to -130\tabularnewline
$p_{e}$ & $\mbox{{[}MeV{]}}$ & 384 & 463 & 509 & 398\tabularnewline
$\theta_{e}$ & $\mbox{{[}deg{]}}$ & 82.4 & 73.8 & 43.4 & 49.4\tabularnewline
$p_{p}$ & $\mbox{{[}MeV{]}}$ & 668 & 495 & 484 & 665\tabularnewline
$\theta_{p}$ & $\mbox{{[}deg{]}}$ & -34.7 & -43.3 & -53.3 & -39.1\tabularnewline
\multicolumn{2}{c}{\# of events} & \multirow{2}{*}{\num{213525}} & \multirow{2}{*}{\num{172142}} & \multirow{2}{*}{\num{2383909}} & \multirow{2}{*}{\num{790365}}\tabularnewline
\multicolumn{2}{c}{after cuts} &  &  &  & \tabularnewline
\end{tabular}
\end{ruledtabular}
\end{table}
In the analysis, cuts were applied to identify coincident electrons and protons that originate from the deuterium target, and to ensure good reconstruction of tracks in the spectrometer and FPP. 
Only events that scatter by more than \ang{10} in the FPP were selected (to remove Coulomb scattering events).
 
Helicity-independent corrections to the measured ratios (acceptance, detector efficiency, target density, etc.) cancel out by the frequent flips of the beam helicity. 
The uncertainties in the beam polarization, carbon analyzing power, and efficiency are also canceled by taking the $P'_x/P'_z$ ratio. 
The total systematic uncertainty in the $P'_x/P'_z$ ratio is estimated to about $1-2\%$ (mainly from reaction vertex reconstruction and spin precession 
estimation) and is small compared  to the dominant statistical uncertainty.

The data for both squared four-momentum transfers are presented in Fig.~\ref{PrlFig2} as a function of the missing-momentum. 
The data are compared to a calculation of $P'_x/P'_z$ for the deuteron~\cite{paper5} that takes into account FSI, MEC, IC and RC.
The calculations presented here (and the hydrogen values used below) use the free proton FFs of Bernauer {\it et al.}~\cite{paper4}. 
The theoretical results shown in Fig.~\ref{PrlFig2} were obtained by averaging calculations on the event-by-event basis over the entire data sample in each bin.
We note that the theoretical results depend on the sign of the missing momentum.
 
Figure~\ref{PrlFig3} shows the double-ratio of the deuteron data to hydrogen, $(P'_x/P'_z)_{^2 \mathrm H}/(P'_x/P'_z)_\mathrm H$, as a function of the {virtuality}.
Our data are supplemented with higher $Q^2$ deuteron data measured at Jefferson Lab~\cite{jlabDeep}. 
The data are shown separately for positive and negative missing momenta to show a possible dependence (as suggested by the calculation).

Our new measurements double the virtuality range covered by the previous experiments. Within the overlap, the data are consistent.
The data show strong deviations of $(P'_x/P'_z)_{^2 \mathrm H}$ from that of a free proton which are indicated by the decrease of the double-ratio well below unity
at large virtuality. The higher-momentum data of JLab, (up to $Q^2$=1.6 $\textrm{GeV}^2$) suggest that the deviation is independent of $Q^2$. 

The deuteron double-ratio data are compared with those of proton knock-out from $^4$He measured at JLab~\cite{paper7}. 
The deuteron is the least bound nucleus in nature with the largest average distance between nucleons and the lowest average nuclear density. 
On the other hand \hefour ~is a very strongly-bound nucleus with a very high average density.

The excellent agreement between the deuteron and \hefour\, data, with the same behavior of the double ratio $(P'_x/P'_z)_A/(P'_x/P'_z)_\mathrm{H}$, Fig.~\ref{PrlFig3}, 
suggests that the deviations from the free proton value due to nuclear effects do not depend on the  nuclear average density. 
It is the virtuality, rather than the nuclear density, that determines the double-ratio behavior, while the latter would 
affect the number of protons of a given virtuality (i.e.\ a given `off-shellness') in the nucleus. 

\begin{widetext}

\begin{figure*}[h!]
\includegraphics[width=0.7\textwidth]{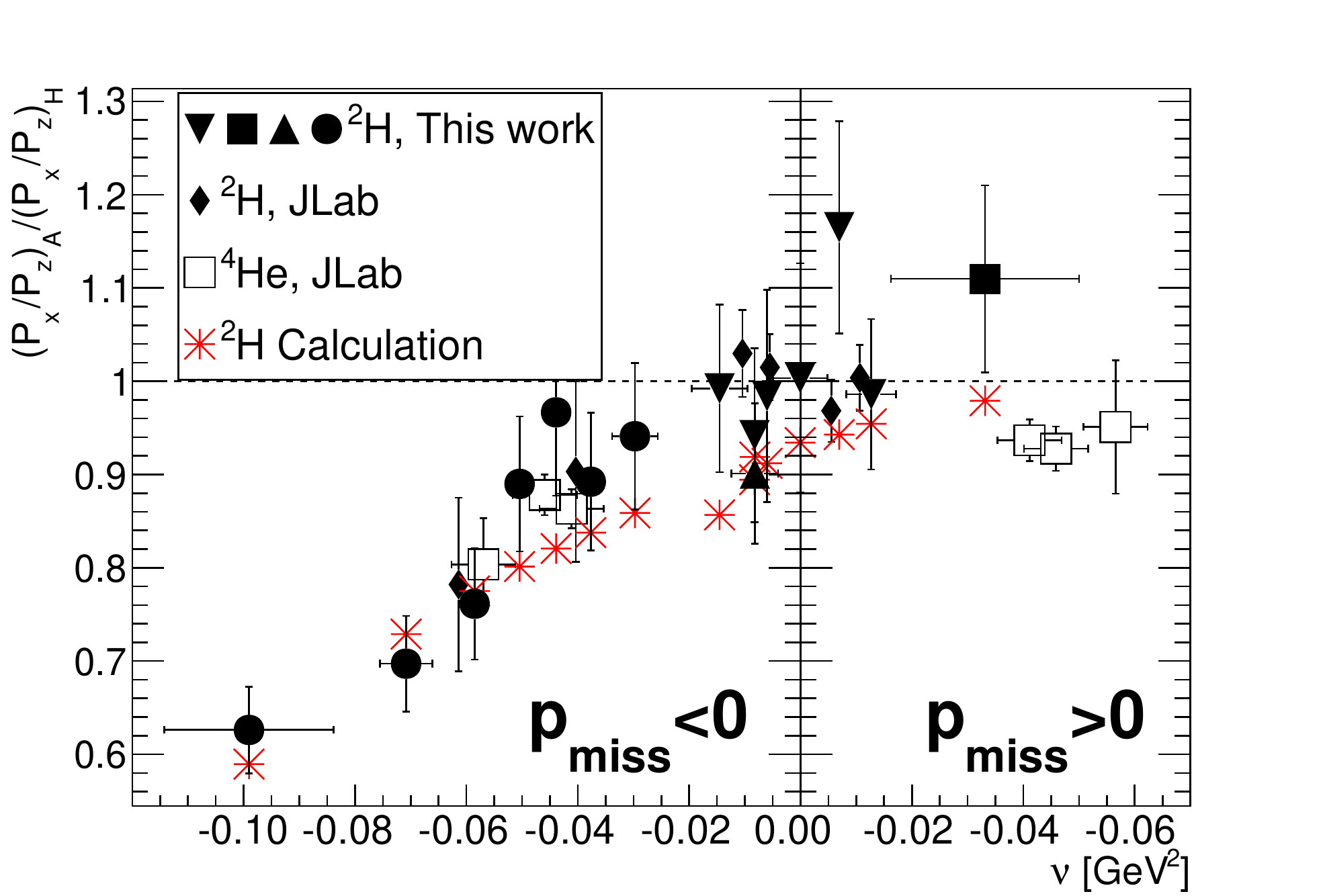}
\caption{\label{PrlFig3}The measured double-ratio $(P'_x/P'_z)_A/(P'_x/P'_z)_H$ ($A= {}^2\mathrm{H},\,^4$He) as a function of the proton virtuality, $\nu$, for deuteron 
(this work and~\cite{jlabDeep}) and for \hefour~\cite{paper7}. 
The virtuality dependence is shown separately for positive and negative missing momenta.
The symbols for the data of this work correspond to those in Fig.~\ref{PrlFig2}.
Also shown are calculations for deuteron case (see text for details).}
\end{figure*}
\clearpage
\end{widetext}

The high precision data reported here allow a meaningful comparison to calculations. 
The calculations~\cite{paper5} consider medium effects but assume {\em unmodified} FFs for the proton in the deuteron. 
They show (see Fig.~\ref{PrlFig3}) a clear virtuality dependence of the ratio $(P'_x/P'_z)_{^2 \mathrm H}$ to that of the free proton which depends on the sign of the missing momentum. 
The contributions of the different corrections in the calculation are not shown in the figure for simplicity. 
However, there is a significant difference between the plane wave impulse approximation and the `full' calculation, mainly due to FSI. 
Relativistic corrections do not alter the slope of the virtuality dependence. 
The calculated contributions of MEC and IC are small.

\begin{figure}
\includegraphics[width=0.535\textwidth]{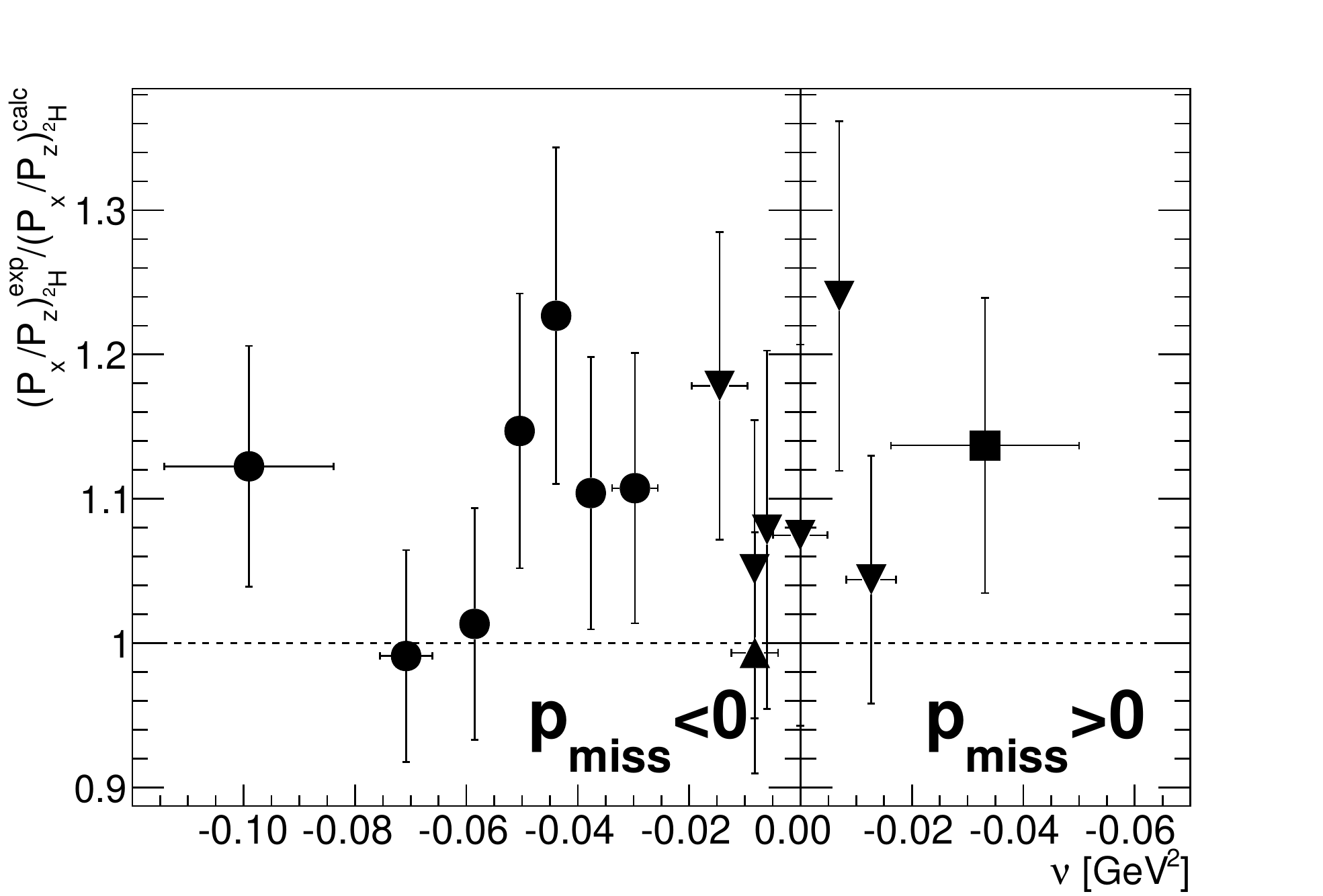}
\caption{\label{PrlFig4}The double-ratio of the measured $(P'_x/P'_z)^{\text{exp}}$ to the theoretical $(P'_x/P'_z)^{\text{calc}}$ (full calculation of~\cite{paper5}). 
The symbols refer to the different experimental set-ups defined in the legend of Fig.~\ref{PrlFig2}.}
\end{figure}

There is an overall deviation of about 10\% between the calculation and the data as shown in Fig.~\ref{PrlFig4}.
It may suggest that additional corrections, such as modifications in the bound nucleon structure may still be required. 
The theoretical results are different for positive and negative $p_{\textrm{miss}}$ kinematics.
This may also be observed also in the $^4$He data~\cite{paper7}. 
This trend should be confirmed for the deuteron with additional data at larger positive $p_{\textrm{miss}}$.

To summarize, the new data of the polarization-transfer double-ratios $(P'_x/P'_z)_{^2 \mathrm H}/(P'_x/P'_z)_H$ extend the previous measurements and almost double the virtuality range. 
The measurements are consistent with the previous $^2$H and $^4$He data sets (obtained in different kinematics). 
They clearly show that the effect of the nuclear medium on $(P'_x/P'_z)_{^2 \mathrm H}$ depend strongly on the virtuality of the bound proton and are almost independent of the average nuclear density and $Q^2$. 
The calculations follow a similar trend as the data and deviate from hydrogen results as the virtuality increases. 
However, taking the form-factors for a bound proton to be those of a free proton, the calculations do not fully reproduce the strong virtuality 
dependence observed in our measurement.
This may indicate the need to invoke in-medium form factor 
modifications. 
It clearly suggests extending the measurements on the deuteron in the positive $p_{\textrm{miss}}$ sector, as well as extending both the $^4$He and deuteron data to larger virtuality. Indeed, such measurements were proposed~\cite{jlabfutureprop} and approved by the Jefferson Lab Program Advisory Committee.
It would be interesting to further examine the effect of virtuality on the proton properties in heavier nuclei.

Notice that in the definition of virtuality Eq.~\eqref{eq_nu}, the full `off-shellness' is associated with the struck proton.
If one assumes equal sharing between the proton and the neutron, none of the essential conclusions of this work will be changed.
However, for heavier nuclei, the way the `off-shellness' will be split between nucleons might yield a large difference.

{
We would like to thank the Mainz Microtron operators and technical crew for the smooth and reliable operation of the accelerator.
This work is supported 
by the Israel Science Foundation (Grant 138/11) of the Israel Academy of Arts and Sciences, 
by the Deutsche Forschungsgemeinschaft (SFB 1044) with the Collaborative Research Center,
and by the U.S. National Science Foundation (PHY-1205782).
}

\end{document}